\documentclass{ws-procs961x669}
\usepackage{float}

\begin{document}


\title{Spontaneous breaking of diffeomorphism invariance in conformally reduced quantum gravity}

\author{G. Giacometti$^{1,2}$\footnote{Speaker}, A. Bonanno $^{3,2}$, S.Plumari$^{1,4}$ and D. Zappalà$^{2,5}$}

\address{$^1$Dipartimento di Fisica e Astronomia “Ettore Majorana”, Università di Catania, Via S. Sofia 64, 95123, Catania, Italy\\ $^2$  INFN, Sezione di Catania, Via Santa Sofia 64, 95123 Catania, Italy\\$^3$ INAF, Osservatorio Astrofisico di Catania, Via S. Sofia 78, 95123 Catania, Italy\\$^4$INFN-Laboratori Nazionali del Sud, Via S. Sofia 62, I-95123 Catania, Italy\\$^5$  Centro Siciliano di Fisica Nucleare e Struttura della Materia, Catania, Italy}

\begin{abstract}
We study the spontaneous breaking of diffeomorphism invariance using the proper-time non-perturbative flow equation in quantum gravity. In particular, we analyze the structure of the UV critical manifold of conformally reduced Einstein-Hilbert theory and observe the occurrence of a non-trivial minimum for the conformal factor at Planckian energies. We argue that our result can be interpreted as the occurrence of a dynamically generated minimal length in quantum gravity.
\end{abstract}


\bodymatter

\section{Introduction}
One of the central challenges in modern theoretical physics is the development of a consistent quantum description of gravity. A crucial aspect of this endeavor is achieving background independence, the notion that the geometric structure of spacetime should not play any role in defining the microscopic degrees of freedom. In classical general relativity, spacetime serves as a fixed background, but in quantum gravity, the background itself must fluctuate and evolve dynamically. Within the framework of the Asymptotic Safety Program \cite{percacci2008asymptoticsafety}, background independence is dynamically realized through the use of the background field method.\\
A useful approximation for studying quantum gravity is conformally reduced gravity, where only the conformal factor of the metric is treated as the dynamical degree of freedom. This approximation offers significant simplifications by reducing the problem to a scalar theory while still capturing essential features of the full gravitational theory. In contrast to classical general relativity, where the conformal factor plays a minor role, in quantum gravity, its fluctuations become dominant in the path-integral formulation due to its inherent instability. These fluctuations are particularly important in both the ultraviolet (UV) regime, near the non-Gaussian fixed point (NGFP) \cite{Hawking:1979ig,PhysRevD.57.971}, and in the infrared (IR) regime, below the Gaussian fixed point (GFP).\\
The key advantage of using conformally reduced gravity is its ability to qualitatively mimic the behavior of the full theory with substantially simpler calculations, owing to its scalar nature. The primary focus of this work is to explore, using the Functional Renormalization Group (FRG) approach, whether a phase transition involving the conformal factor can occur at low energies, leading to a non-zero vacuum expectation value $\left<\chi\right>\ne0$. Such a phase transition could signal a phase of broken diffeomorphism invariance, where the metric $\left<g_{\mu\nu}\right>\ne0$ emerges as a low-energy phenomenon. One of the first works showing this kind of behavior has been done by Reuter and Weyer\cite{Reuter_2009}, using the Wetterich equation only for the effective potential to find, numerically, possible solutions. \\ 
Another study, still in the Local Potential Approximation (LPA), has been done by Bonanno and Guarnieri \cite{Bonanno:2012dg}, with the proper time equation. We aim to extend this analysis by incorporating the evolution of the $Z_k$ term in the proper time framework. This inclusion will provide a more comprehensive understanding of the dynamical behavior of the conformal factor and its role in the emergence of spacetime at low energies.\\
This approach has potential applications in cosmological models, particularly in scenarios such as inflation, where the emergence of spacetime at specific energy scales could play a critical role in early-universe dynamic\cite{Cecchini_2024,Rinaldi_2016}.

\section{Effective Action for the conformal factor}
In this section, we present the basics of the Wilsonian Functional Renormalization Group (FRG) approach and explain how it can be applied to the conformal factor. We begin with the action $\mathcal{S}$ for the fundamental field $\chi(x)$ which can be decomposed as $\chi(x)=\chi_B(x)+f(x)$ where $\chi_B(x)$ represents a static background field, and $f(x)$ denotes a fluctuating field. The background metric $\hat{g}_{\mu\nu}$ refers to a rigid reference metric defined on a Euclidean $d$-dimensional manifold.\\
The Wilsonian action is defined as:
    \begin{equation}
        e^{-S_k[\Tilde{f};\chi_B]}=\int D[f]\delta(f_k-\Tilde{f})e^{-\mathcal{S}[\chi_B+f]}
        \label{eq:wilson}
    \end{equation}
where $f_k(x)$ is an averaged fluctuation field given by:
    \begin{equation}
        f_k(x)=\int d^dy\sqrt{\hat{g}}f(y)\rho_k(y,x;\chi_B)
    \end{equation}
with $\rho_k(y,x;\chi_B)$ as the smearing kernel.\\
In the full theory, a background metric $\bar{g}_{\mu\nu}$ is chosen to perform the calculations, while the fluctuations $h_{\mu\nu}$ are quantized around this background, subject to the condition$\left<h_{\mu\nu}\right>\equiv\bar{h}_{\mu\nu}=0$. In the conformally reduced gravity approximation, the analogous relations are:
    $$\bar{f}\equiv\left<f\right>\longleftrightarrow\bar{h}_{\mu\nu}\equiv\left<h_{\mu\nu}\right>$$
    $$\phi\equiv\left<\chi\right>=\chi_B+\bar{f}\longleftrightarrow g_{\mu\nu}\equiv\left<\gamma_{\mu\nu}\right>=\bar{g}_{\mu\nu}+\bar{h}_{\mu\nu}$$
The background metric is chosen as:
    \begin{equation}
        \bar{g}_{\mu\nu}\equiv\chi_B^{2\nu}\hat{g}_{\mu\nu}
    \end{equation}
where $\nu\equiv2/(d-2)$ and $d$ is the space-time dimension.\\
The fixed metric is used to perform the calculations, while all dynamical fields are spectrally decomposed using the eigenbasis of $-\bar{\Box}$, where the eigenvalues satisfy:
    \begin{equation}
        \bar{k}^2=\chi_B^{-2\nu}k^2
    \end{equation}
For $k=0$ the blocked action (\ref{eq:wilson}) coincides with the effective potential. For $k\ne0$ it acts as an effective action for the modes with momentum $p<k$. The relation to the standard effective average action for the conformal factor \cite{Reuter:2008wj} can be derived by recognizing that the expectation value of the blocking field is given by:
    \begin{equation}
        \left<\Tilde{f}\right>=\frac{1}{\sqrt{\hat{g}}}\left.\frac{\partial W_k[J;\chi_B}{\partial J(x)}\right|_{J=0}
    \end{equation}
with the generating functional $W_k[J;\chi_B]$ defined as
    \begin{equation}
        W_k[J;\chi_B]=\int d^dx\{J(x)\Tilde{f}(x)-S_k[\Tilde{f};\chi_B]\}
        \label{eq:effact}
    \end{equation}
This equation represents the standard generating functional \cite{Reuter:2009kq}.\\ 
Using the proper-time formalism, we can derive a flow equation that yields a regularized one-loop contribution as follows:
    \begin{equation}
        k\partial_kS_k[\Tilde{f};\chi_B]=-\frac{1}{2}\text{Tr}\int_0^\infty\frac{ds}{s}k\partial_k\tau_k\text{exp}\left\{-s\frac{\delta^2S_k[\Tilde{f};\chi_B]}{\delta\Tilde{f}^2}\right\}
    \end{equation}
where the one-parameter smooth cutoff function $\tau_k\equiv\tau_k^n$ is given by\cite{multi35}
    \begin{equation}
        \tau_k^n=\frac{\Gamma(n,s\mathcal{Z}nk^2\chi_B^{2\nu})-\Gamma(n,s\mathcal{Z}n\Lambda_{\text{cutoff}}^2\chi_B^{2\nu})}{\Gamma(n)}
    \end{equation}
The derivative with respect to $k$ is
    \begin{equation}
        k\partial_k\tau_k^n(s)=-\frac{2}{n!}(s\mathcal{Z}nk^2\chi_B^{2\nu})^n\text{exp}(s\mathcal{Z}nk^2\chi_B^{2\nu})
    \end{equation}
with the constrain that $n>d/2$. In the limit $n\rightarrow\infty$, this corresponds to the sharp-cutoff\cite{Floreanini_1995}.\\
It is important to emphasize that in this work, a particular topology is used merely as a technical tool to project the infinite-dimensional flow equation into a finite-dimensional theory space. However, the functional flow equation is, by construction, independent of the topology \cite{PhysRevD.57.971}, and this property is preserved for the flow of the conformal factor.\\
Our starting point is the Einstein-Hilbert action\cite{Reuter:2008wj,Reuter:2009kq}
\begin{equation}
    S_k^{EH}[g_{\mu\nu}]=-\frac{1}{16\pi}\int d^dx\sqrt{g}G^{-1}_k(R(g)-2\Lambda_k)
\end{equation}
After performing a Weyl rescaling, the action becomes:
\begin{equation}
    S_k[\phi]=\int d^dx\sqrt{\hat{g}}Z_k\left(\frac{1}{2}\hat{g}^{\mu\nu}\partial_\mu\phi\partial_\nu\phi+\frac{1}{2}A(d)\hat{R}\phi^2-2 A(d)\Lambda_k\phi^{\frac{2d}{d-2}}\right)
\end{equation}
where $\hat{R}\equiv R(\hat{g})$, and
    \begin{equation}
        Z_k=-\frac{1}{2\pi G_k}\frac{d-1}{d-2},\quad A(d)=\frac{d-2}{8(d-1)}
        \label{eq:relation}
    \end{equation}

\subsection{$\mathbb{R}^4$ Projection}
When we project on a flat topology, such as $\mathbb{R}^4$, the quadratic term of the Weyl action vanishes, leading to an action of the form:
    \begin{equation}
        S_k[\phi]=\int d^4x\sqrt{\hat{g}}\left(\frac{1}{2}Z_k\hat{g}^{\mu\nu}\partial_\mu\phi\partial_\nu\phi+V_k(\phi)\right)  
    \end{equation}
where $V_k(\phi)=Z_kU_k(\phi)$ represents a generic potential that we will define later.\\
Our goal is to compute two coupled flow equations for both $V_k(\phi)$ and $Z_k$. Using the derivative expansion around a fixed background and considering small fluctuations, we define the field as:
$$\phi=\chi_B+\Tilde{f}(x)$$
After performing the calculations, the action becomes:
    \begin{multline}
        S_k[\phi]=\int d^4x\sqrt{\hat{g}}\Bigg(-\frac{1}{2}Z_k\Tilde{f}(x)\hat{\Box}\Tilde{f}(x)+\frac{1}{2}V''_k[\chi_B]\Tilde{f}(x)^2+\\\frac{1}{3!}V'''_k[\chi_B]\Tilde{f}(x)^3+\frac{1}{4!}V^{(IV)}_k[\chi_B]\Tilde{f}(x)^4\Bigg)  
    \end{multline}
where we have only included terms that will contribute after taking the second derivative.\\
Next, we write the flow equation as:
    \begin{equation}
        k\partial_kS_k[\Tilde{f}(x)]=-\frac{1}{2}\int d^4x\sqrt{\hat{g}}\chi_B^{4}\int\frac{ds}{s}k\partial_k\tau_k^n\left<x|e^{-s(K+\delta K)}|x\right>
    \end{equation}
Where $K$ and $\delta K$ are defined as:
$$K=-Z_k\hat{\Box}+V''_k[\chi_B]$$
$$\delta K=V'''_k[\chi_B]\Tilde{f}(x)+\frac{1}{2}V^{(IV)}_k[\chi_B]\Tilde{f}(x)^2$$

To evaluate the trace in a background-independent manner, we integrate over the eigenvalues of the Laplacian constructed from the background metric $\bar{g}_{\mu\nu}$, using the identity $\int d^d\bar{p}\left|\bar{p}\right>\left<\bar{p}\right|=\mathbb{I}(2\pi)^d$ and substituting $\hat{\Box}\rightarrow\bar{\Box}\chi_B^2$. By applying the Baker-Campbell-Hausdorff (BCH) expansion, we obtain:
\begin{multline}
    k\partial_kS_k[\Tilde{f}(x)]=-\frac{1}{2}\int d^4x\sqrt{\hat{g}}\chi_B^4\int \frac{d^4\bar{p}}{(2\pi)^4}\int\frac{ds}{s}\\k\partial_k\tau_k^n\left<x\right|\bar{p}\left>\right<\bar{p}|e^{-sK}(1-s\delta K+\frac{s^2}{2}\{[\delta K,K]+\delta K^2\}\left|x\right>
\end{multline}
After lengthy calculations, we obtain the desired set of coupled partial differential equations (PDEs):
    \begin{equation}
        \!
        \begin{aligned}
            k\partial_kV_k=M(k^2\chi_B^2)^{2}\frac{1}{\left(1+\frac{V"_k(\chi_B)}{k^2nZ_k\chi_B^2}\right)^{n-2}} \\
            k\partial_kZ_k=N(k^2\chi_B^2)^{-1}\frac{(V'''_k/Z_k)^2}{\left(1+\frac{V"_k(\chi_B)}{k^2nZ_k\chi_B^2}\right)^{n+1}} 
        \end{aligned}\label{eq:set}
    \end{equation}
where the coefficients $M$ and $N$ are given by:
$$M=\left(\frac{n}{4\pi}\right)^2\frac{\Gamma(n-2)}{\Gamma(n)}$$
$$N=\frac{(4 - 2 (n + 1))(4 - 2 (n + 2))}{96n^2}\left(\frac{n}{4\pi}\right)^2\frac{\Gamma(n-2)}{\Gamma(n)}$$
In the results discussed in the next section, the value of the parameter $n$ will be settled equal to 3.\\
A more complex system of equations has already been derived in previous work \cite{bonanno2023conformalsectorquantumeinstein}, where other truncations of the theory have been studied. However, a full numerical evaluation of these systems has not yet been completed due to the complexity of the equations involved.

\section{Numerical Analysis}
We now turn to the numerical calculations and comment on our findings. First, it is important to note that in the system described by Eqs. (\ref{eq:set}), the field $Z_k$ evolves as a constant and does not depend explicitly on the field $\chi$. We work in the so-called LPA' approximation, where $Z_k$ is evaluated based on the value of the minimum of $V_k(\chi)$ through time. The position of the minimum is evaluated step by step by solving both the PDEs and numerically searchin it. Our initial conditions for the potential $V_k(\chi)$ and the function $Z_k$ are as follows:
$$V(\chi,k_0)=\frac{\lambda}{6}\chi^4+\sigma\chi^6+\omega\chi^8$$
$$Z(k_0)=c(k_0)$$
Where $Z_k$ is treated as a constant that depends only on the initial scale, using the value of the dimensionfull Newton constant $G=1$ in the relations \ref{eq:relation}.\\
Let us first briefly discuss the initial potential, the two extra terms, $\sigma\chi^6$ and $\omega\chi^8$, arise from higher powers of volume operator of the form $\mathcal{V}=\int d^dx \sqrt{g}$, which are necessary to drive the transition to a phase of broken diffeomorphism invariance in the infrared (IR) region.\\
Before proceeding with the numerical results, it is essential to clarify the renormalization procedure. In typical FRG calculations, the flow starts from an ultraviolet (UV) initial condition, integrating towards the IR region. This is done using the standard definition of the renormalization time:
$$t=-\ln(k/\Lambda)$$
However, this cannot be applied directly in our case, as $Z<0$ leads to the evolution equation becoming a backward-parabolic equation, a class of diffusion-like partial differential equations (PDEs) with a negative diffusion constant. To establish a well-posed Cauchy problem, we must reverse the flow direction—from IR to UV—by defining time as $k\equiv\Lambda e^t$ with $t>0$. This allows for the evolution towards the UV regime. Once the deep UV solution is obtained, it can be treated as admissible initial data for a non-singular IR flow that ultimately aligns with our ansatz.

\subsection{Evolution of the Potential}
\begin{figure}[H]
    \centering
    \begin{minipage}{0.48\textwidth}
        \centering
        \includegraphics[width=\linewidth]{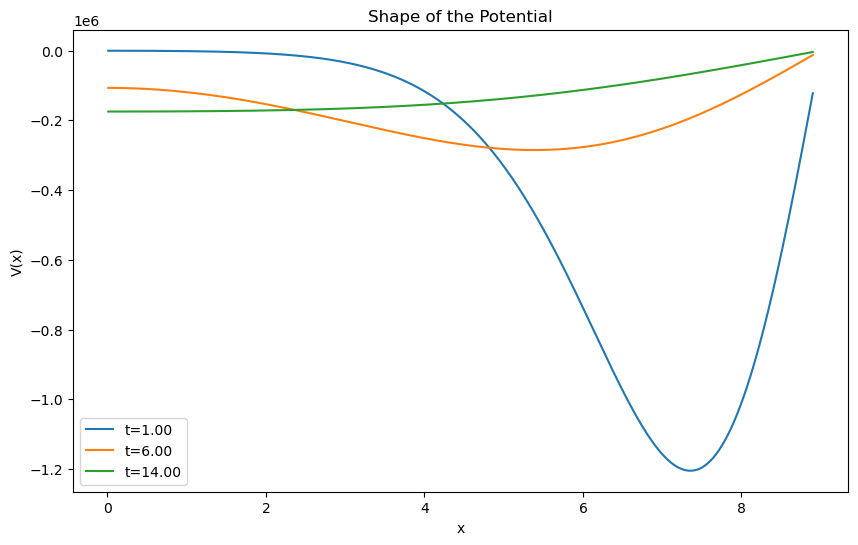}

    \end{minipage}
    \hfill
    \begin{minipage}{0.48\textwidth}
        \centering
        \includegraphics[width=\linewidth]{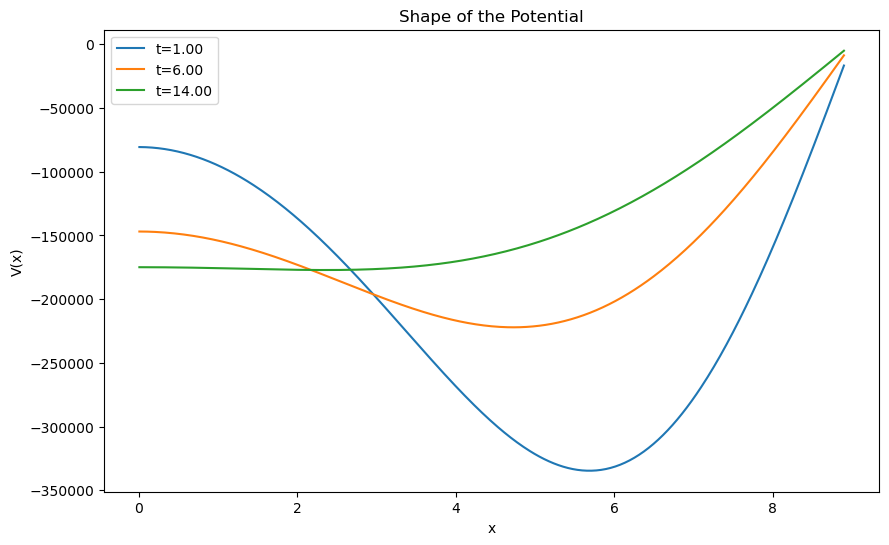}

    \end{minipage}

    \vspace{0.5cm} 

    \begin{minipage}{0.48\textwidth}
        \centering
        \includegraphics[width=\linewidth]{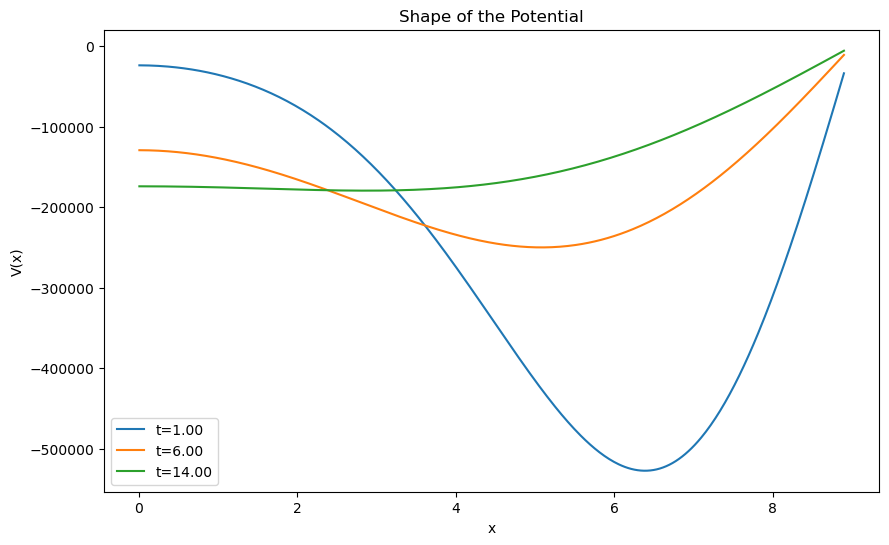}

    \end{minipage}
    \hfill
    \begin{minipage}{0.48\textwidth}
        \centering
        \includegraphics[width=\linewidth]{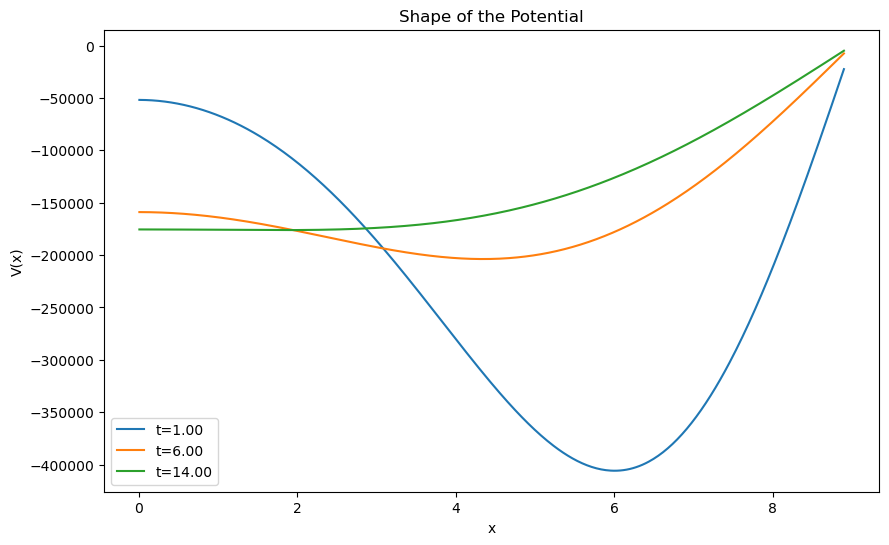}

    \end{minipage}
    \caption{Evolution of the potential for different values of the coefficients in the initial condition}
    \label{fig:potential_images}
\end{figure}

With the foundations of our numerical framework established, we set $n=3$ as the proper-time parameter for the calculation. We begin by examining the evolution of the potential for different initial conditions as shown in figure \ref{fig:potential_images}
In these figures, we observe that the potential starts with a non-trivial minimum in the IR region (the early stage of the evolution) and gradually approaches a trivial minimum as we move towards the UV. This behavior is consistent with the expectation that in the UV regime, the system recovers the trivial position of the minimum. Previous efforts to reconstruct this type of potential have been made through the use of beta functions within the Einstein-Hilbert truncation \cite{Daum:2009qe}, where the correct behavior was inferred. However, our approach follows the full evolution of the potential, providing a more detailed picture.

\subsection{Evolution of $Z_k$}
\begin{figure}[H]
        \centering
        \includegraphics[width=0.6\textwidth]{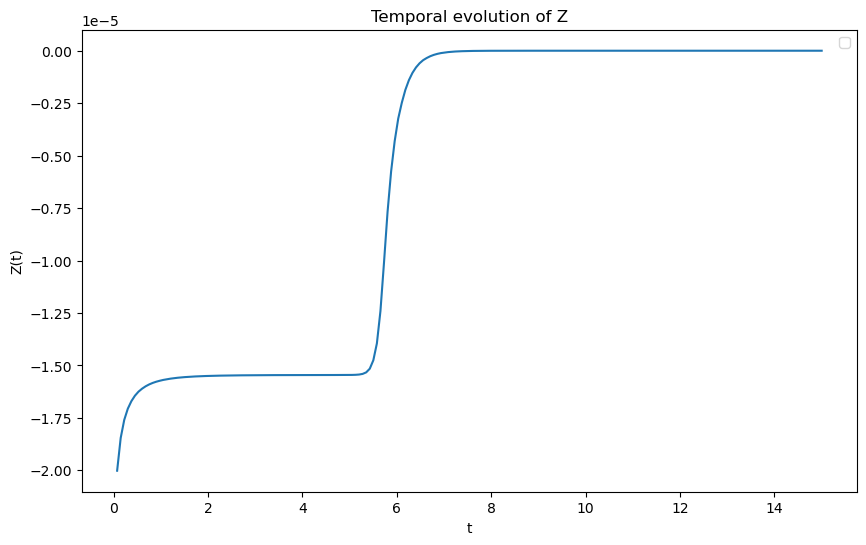}
        \caption{Time evolution of Z}
        \label{fig:graficoz}
    \end{figure}

In addition to the potential, we also track the evolution of $Z_k$ using the flow equation. The results are presented in figure \ref{fig:graficoz}
The evolution of $Z_k$ is directly connected to the evolution of the dimensionless Newton constant $G$, through the relation given in Eq. (\ref{eq:relation}). The plot shows that $Z_k$ eventually reaches a plateau, small but non-zero, indicating the approach to the non-Gaussian fixed point (NGFP) in the UV regime. This is an important result as it provides insight into the behavior of the system at high energies.

\subsection{Evolution of the Minimum Position}
\begin{figure}[H]
    \centering
    \includegraphics[width=0.6\textwidth]{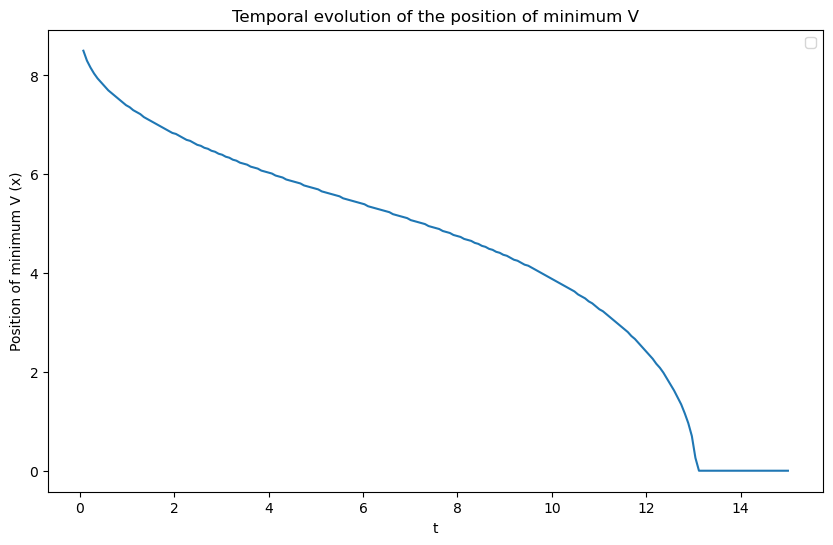}
    \caption{Evolution through time of the minimum position, at t=13.04 we reach the zero value}
    \label{fig:immagine}
    \end{figure}

Finally, we focus on the main result of this study: the evolution of the position of the potential minimum over time that is presented in figure \ref{fig:immagine}
The figure clearly illustrates a smooth evolution from the initial non-trivial minimum to the trivial minimum at later times. A critical transition occurs around $t=13.04$, where the minimum reaches the zero value. This transition time is of particular interest for our physical analysis, as it marks the point where the potential enters the phase of broken diffeomorphism invariance.

\subsection{Physical analysis}
Our primary interest lies in the generation of a "minimal length" that allows us to distinguish between a phase of broken diffeomorphism invariance and an unbroken one. Previous work has established \cite{Reuter_2007} that we can relate the expectation value of the metric to both the conformal factor and the scale at which we probe spacetime. These relations are expressed as:

$$\left<g_{\mu\nu}\right>=\left<\phi^2\right>\hat{g}_{\mu\nu}$$

and

$$\left<g_{\mu\nu}\right>=l(k)\hat{g}_{\mu\nu}$$
where $l(k)$ represents the characteristic length scale at energy scale $k$. These two expressions allow us to connect the position of the potential minimum with the length scale at each energy level.\\
The minimal length corresponds to the scale at which the position of the potential minimum reaches zero. This can be evaluated using the relation:
    \begin{equation}
        l(k_{\text{tr}})=\frac{\pi}{k_\text{tr}}=\frac{\pi e^{-t}}{\Lambda_\text{tr}}
    \end{equation}
where $k_\text{tr}$ represents the transition scale, and $\Lambda_\text{tr}$ is the UV cutoff. With the result from our example at $t=13.04$, we obtain:
    \begin{equation}
         l(k_{\text{tr}})\approx 7.10\cdot10^{-6} l_{\text{IR}}
    \end{equation}
where $l_\text{IR}$ is the infrared length scale, i.e. the scale at which the flow goes near the Gaussian fixed point.  We can set this IR length to the Planck length, $l_\text{P}=1.16\cdot10^{-33}\text{cm}$, as it represents the smallest scale where quantum gravitational effects are not required. This leads to the minimal length estimate:

    \begin{equation}
         l(k_{\text{tr}})\approx 1.13\cdot10^{-38}\text{cm}
    \end{equation}

To assess the robustness of these results, we tested different initial values for the coefficients of the potential. The majority of our starting potentials exhibited a transition time in the range $12<t<14$, resulting in minimal length values in the range:

    \begin{equation}
        4\cdot10^{-39}\text{cm}\le l(k_{\text{tr}})\le3\cdot10^{-38}\text{cm}
    \end{equation}

A more thorough statistical investigation is necessary to solidify these findings, which would include a more extensive set of simulations. Additionally, an exploration of how varying the cutoff parameter $n$ influences the results is required to enhance the reliability of the minimal length estimate.

\section{Conclusions}
In this work, we have investigated the flow equation of conformally reduced gravity beyond the Einstein-Hilbert truncation, aiming to identify a transition scale associated with the generation of spacetime. By incorporating higher-order powers of the volume term, we explored infrared (IR) potentials exhibiting a non-trivial minimum. Through varying the coefficients, we obtained a range of values that could correspond to a minimum length scale below which the concept of geometry ceases to exist.\\
We examined the LPA' evolution of the equation for $Z_k$ in the proper time framework, we observed that this minimum length scale emerges in conjunction with the Newton constant reaching its fixed-point value. This result shows similar behavior to the results of Reuter and Weyer, obtained with a different FRG equation, indicating the validity of the result. The reliability of the numbers obtained hinges on two critical considerations. First, we have worked within the conformally reduced approximation, which provides a qualitative insight into the evolution of the full theory, although it may fail to reproduce the exact numerical values. As a result, the minimal length scale could differ if a similar investigation were conducted using the full gravitational theory. Second, we considered only the gravitational field, without the inclusion of matter fields, which are crucial for describing spacetime geometry in this context. The presence of coupled matter fields may significantly alter the results.\\
In summary, our findings suggest that in quantum gravity, spacetime generation may occur at a specific energy scale—and thus at a corresponding length scale—implying that spacetime could be an emergent phenomenon in the low-energy limit. This work will be further pursued to go beyond the LPA', by solving an equation for a field dependent $Z_k(\chi)$.

\bibliographystyle{ws-procs961x669}
\bibliography{ws-pro-sample}

\end{document}